\documentclass[amssymb,amsmath,prd,twocolumn,floatfix,showpacs]{revtex4}
\usepackage{graphicx}
\usepackage{lastpage}
\usepackage[normalem]{ulem}
\usepackage{epstopdf}
\graphicspath{ {Images/} }

\begin{document}
\title{Energetics of the Quantum Graphity Universe}

\author{Samuel A.~Wilkinson}
\affiliation{Chemical and Quantum Physics, School of Applied Sciences, RMIT University, Melbourne 3001, Australia}

\author{Andrew D.~Greentree}
\affiliation{Chemical and Quantum Physics, School of Applied Sciences, RMIT University, Melbourne 3001, Australia}
\email{andrew.greentree@rmit.edu.au}

\date{\today}

\begin{abstract} Quantum graphity is a background independent model for emergent geometry, in which space is represented as a dynamical graph. The high-energy pre-geometric starting point of the model is usually considered to be the complete graph, however we also consider the empty graph as a candidate pre-geometric state. The energetics as the graph evolves from either of these high-energy states to a low-energy geometric state is investigated as a function of the number of edges in the graph. Analytic results for the slope of this energy curve in the high-energy domain are derived, and the energy curve is determined exactly for small number of vertices $N$. To study the whole energy curve for larger (but still finite) $N$, an epitaxial approximation is introduced. This work may open the way  to compare predictions from quantum graphity with observations of the early universe, making the model falsifiable.

\end{abstract}


\maketitle

\section{Introduction}

Quantum graphity (QG) is an intriguing and speculative idea to explain how our familiar concepts of space-time might arise from some more fundamental considerations \cite{bib:KMS2008}.  In particular, QG models consider the universe to be initially in some high energy state with no notion of geometry.  This state is allowed to relax under some Hamiltonian to a low energy state in which geometry is emergent. The onset of familiar geometric space, \emph{geometrogenesis}, is a central point of focus of quantum graphity. 

Traditionally, the pre-geometric starting point of the model has been taken to be the complete graph on $N$ vertices, $K_N$, in which every vertex is connected to every other. When space is represented by a complete graph, concepts of locality, distance and direction are at best poorly defined, and certainly cannot explain the Universe as it is observed today.  It is therefore necessary to posit mechanisms by which the graph can relax to some global minimum, which is presumed to have the properties of geometry that we experience.  Mechanisms to allow such relaxation include the formation of matter on the graph \cite{bib:Hamma2010}, and equilibration with some external heat bath \cite{bib:Cara2011}.

The use of graphs to represent spacetime is common in background independent models, such as loop quantum gravity \cite{bib:Rovelli2004}, spin foam models \cite{bib:Perez2013}, and quantum causal histories \cite{bib:Mark2000} due to the fact that graphs are purely combinatorial objects and therefore make no reference to an inherent background geometry. Of particular relevance to QG is causal dynamical triangulations \cite{bib:Ambj2006}. These two models share many features, in particular the notion of geometry arising as an emergent feature of a model that does not assume it, simulatability \cite{bib:Loll2008} and the existence of a geometrogenic phase transition \cite{bib:Miel2014}. QG also has many similarities with a recently emerging class of condensed matter analogue models for spacetime \cite{bib:Sind2012}, which include models in which a quantum theory of gravity emerges as an effective field theory of some more fundamental structure \cite{bib:Bain2013}, such as the approach of Wen and collaborators in which treating the vacuum as a bosonic spin system leads to the existence of photons, electrons and gravitons as low-lying excitations \cite{bib:Gu2012}, and the concept that spacetime itself may be a Bose-Einstein condensate \cite{bib:Hu2005}. QG fits this category, as it takes seriously and literally the notion that spacetime may be a sort of condensed matter system, and can be treated using the concepts and techniques of statistical physics and many-body theory. The key difference between QG and other theories of emergent spacetime from a discretized model is that in QG the lattice itself is dynamical rather than fixed.

QG has been shown to give rise to primitive gravitational behaviour, as seen in a toy model of a black hole \cite{bib:Cara2012}. When the model includes additional degrees of freedom on the edges, the ground state of the model is a string-net condensate \cite{bib:KMS2008,bib:LW2005}, from which photons and electrons may emerge as low-lying excitations \cite{bib:LevWen2005}. Lieb-Robinson bounds for such a system have been derived, leading to an emergent speed of light in this model \cite{bib:HMPS2009}. See \cite{bib:Hamma2011} for a review.

Here we seek to understand some of the properties of how the Universe might evolve under conditions of QG.  In particular, we calculate the energy of states at or close to the QG ground state by explicitly calculating the energy of the QG graph as a function the number of edges for certain finite cases, and in the infinite limit.  In addition to the complete graph as being a candidate initial state for the Universe, our results highlight the fact that the empty graph, i.e. the graph with $N$ vertices and no edges, is a candidate pre-geometric graph. Because the rate of change of the Hamiltonian with respect to total edge number is different in the limits that the number of edges tends to infinity, or tends to zero, future work may identify observational means to distinguish these two possibilities. Figure~\ref{fig:IconicFigure1} shows a schematic of the three main regions of interest: the two possible starting points (the complete graph and the empty graph)  and the low energy geometric state (here assumed to be a honeycomb lattice), with a rough sketch of the energy curve connecting these three regions.

\begin{figure}
\centering
\includegraphics[width=1\linewidth]{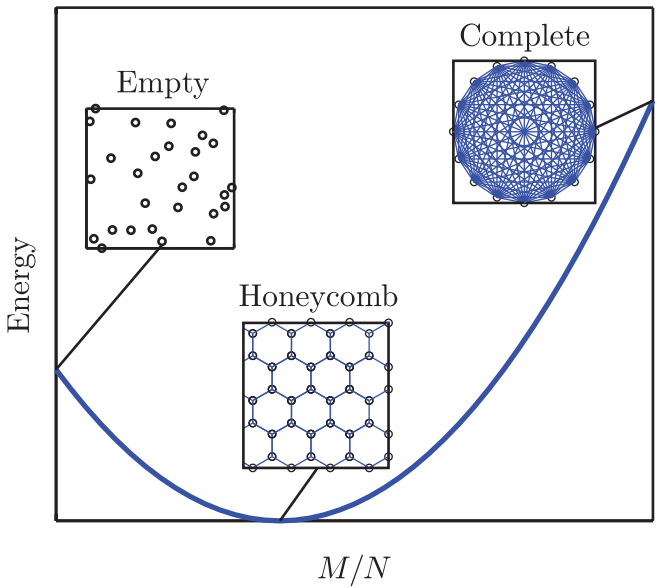}
\caption{\label{fig:IconicFigure1} Schematic of the three regions of interest of the graph and the the energy curve connecting them. This includes two possible high-energy pre-geometric graphs: the empty graph that consists of disconnected vertices with no edges, and the complete graph where every vertex is connected to every other. Each of these high energy graphs can lower their energy by created/deleting edges, as thus tend towards the low-energy geometric graph, here represented as a honeycomb lattice.}
\label{fig:IconicFigure1}
\end{figure}

It is important to stress that determining the global ground state of the QG model, and indeed determining any local ground state as a function of the number of edges, is equivalent to graph isomorphism: a problem for which there exists no known P algorithm \cite{bib:Garey1983}.  Hence it is necessary to make certain assumptions about the likely evolution of states, and practical considerations such as metastability and frustration are likely to play a significant role in the `true' evolution of the Universe, which is likely to lead to defects in the emergent graph \cite{bib:QSM+2012}. To calculate the energetics of the graph, we consider variations accessible by the addition or deletion of a single edge.  We term this constraint the epitaxial approximation.

This paper is organised as follows.  We first introduce the QG Hamiltonian and the particular parameter set that we are using to analyse it.  We then discuss the total energetics of the QG Hamiltonian as a function of the number of edges, under the epitaxial approximation.  Finally we offer speculation for possible observational consequences of these results so as to suggest tests to falsify the QG model.

\section{The Model}

To construct a QG model in which geometry is an emergent coarse-grained property of space, we represent space by an abstract graph $G$ with no \emph{a priori} notion of geometry. A graph consists of a set of vertices $\mathbb{V} = \{\nu_i\}$ and set of edges, $\mathbb{E} = \{(\nu_i,\nu_j)\}$, where the $\nu_i$  correspond to points in space, and the edges correspond to adjacencies between these points [i.e. two vertices $\nu_i$ and $\nu_j$ can be considered adjacent if the edge $(\nu_i,\nu_j) \in \mathbb{E}$].

Following Konopka \textit{et al.}, we apply the canonical QG Hamiltonian to the graph $G$ 
\begin{align}
H = H_V + H_L + H_{\textrm{hop}}, \label{eq:Ham}
\end{align}
where $H_V$ is the valence  term, describing the number of edges per vertex; $H_L$ the loop term, counting the size of plaquettes; and $H_{\textrm{hop}}$ the hopping term, which describes the motion of edges on the graph.  Explicitly, the valence term is
\begin{align}
H_V = g_V\sum\limits_i e^{p(v_i-v_0)^2}
\end{align}
where $v_i$ is the valence, or \emph{degree}  of vertex $\nu_i$, $v_0$ is the ideal valence of the graph, $p$ is a dimensionless real number and $g_V$ is a positive coupling constant. This term favours regular graphs where every vertex has degree $v_0$.

The loop term $H_L$  is
\begin{align}
H_L = -g_L\sum_{L=3}^{L_{\max}}\frac{r^L}{L!}\sum_a P(a,L),
\end{align}
where $P(a,L)$ is a function that counts the number of loops of length $L$ that pass through vertex $a$, $r$ is a dimensionless real number and $g_L$ is a positive coupling constant. For our purposes, a loop is defined as a path that begins and ends at the same vertex and that includes no vertex more than once (except for the initial/final vertex). The sum over loop lengths $L$ begins at 3, because that is the length of the shortest possible non-retracting loop. Ideally, the sum would extend upwards to include loops of infinite length, but to the make the model computationally tractable, loop counting is truncated at some maximum length $L_{\max}$. The weighting factor $r^L/L!$ is small both when $L$ is small and when $L$ is large. Between these points it reaches a peak at some value $L_0$, which is determined by $r$. Thus arbitrarily long loops contribute a negligible amount of energy, justifying the use of a truncation length $L_{\max}$, and by varying $r$ we can tune the Hamiltonian so that loops of some desired length $L^*$ contribute most. The negative sign in front means that this term lowers the energy of the graph, so that it favours graphs with predominantly loops of length $L^*$.

The last term, $H_{\textrm{hop}}$, is a kinetic term that allows edges to hop about the graph, changing the overall configuration. The exact form of this term is not relevant for our purposes, although the presence of such a term is necessary so that the graph may dynamically evolve without changing the number of edges.

In the Hamiltonian of eq.~\ref{eq:Ham}, the total number of edges is a constant of the motion.  Edge creation or deletion is therefore a non-energy conserving process.  As mentioned above, this implies some form for thermal reservoir is necessary to force evolution to the ground state \cite{bib:Cara2011}, which may be problematic as the graph represents the entire Universe. This problem is addressed by associating the loss of energy of the graph with the creation of matter (and vice versa, the annihilation of matter can raise the energy of the graph) \cite{bib:Hamma2010}.  Nevertheless, we do not address such questions here, instead concentrating on the energetics of the approach of the Universe to the ground state as a function of the number of edges.

The loops term in the Hamiltonian presents significant computational difficulty. For graphs on small $N$, a backtracking algorithm based on the method of Franzblau \cite{bib:Franz1991} is employed. For large $N$ this method becomes computationally prohibitive, so instead a method using explicit formulas for the number of loops of a given length is used (discussed below). This more direct approach is an approximation, due to the fact that explicit formulas for loops of $L>7$ are not known \cite{bib:Pere2014}. The fact that longer loops do not contribute in this method means that in general, extended lattices are not favourable (with no contribution due to loops of length $>7$ there is no energetic benefit to the formation of connected structures on more than 7 vertices).

Following Konopka \textit{et al.}, the values of the Hamiltonian parameters chosen are $p = 1.2$, $r = 6.5$, $g_V/g_L~=~500$, and  $v_0 = 3$. The choice of parameters must, within some limits, determine the resultant ground state of the QG Universe, the rate of approach of some initial state to the ground state, and the rate at which matter is created on the graph.  Hence it is important to understand what observational consequences there are to these choices, and whether it is possible to constrain these parameters based on observations of the Universe and the Big Bang.  These parameters are chosen to impose a regular 2-D honeycomb graph as the ground state, although it is not known if these parameters favour a honeycomb graph as the true ground state of the system, or if the honeycomb graph is a local minimum. For computation tractability, loop counting is truncated at $L_{\max} = 14$ in the algorithmic approach and $L_{\max} = 7$ in the explicit approach. At present, these parameters can only be used to generate insight into the properties of QG, and should not be considered as explicitly describing the Universe in which we live.

\section{Energetics of Geometrogenesis}
\subsection{Analytic results}

The evolution of the QG universe is postulated to begin in a high-energy pre-geometric state.  This pre-geometric state is typically considered as the complete graph, but here we also discuss the  empty graph. The Universe then proceeds towards some lower energy state.  Because the number of edges of the ground state is different from either of the two likely starting states, this evolution must perforce include either or both edge creation and deletion, as well as reorganisation of the graph to the local minimum with a given ratio of number of edges to number of vertices. Considering both creation and deletion of edges naturally suggests the concept of \textit{holes} in the QG model, and we use the concept of holes here to refer to the absence of an edge.  We will show that the QG Hamiltonian described above shows marked particle-hole asymmetry.

The rate of edge creation and deletion compared with the rate of edge hopping is not known, so we shall consider this as a two-step process. First, we will assume that a single edge is created/deleted; then we allow the graph sufficient time for every edge to hop until the graph energy reaches the minimum energy for this number of edges before another edge is created/deleted.   Therefore each step along the path from the empty or complete graph to a low-energy ground state graph corresponds to a ground-state graph for a fixed number of edges, $M$, where the number of holes is equivalently $M^* = [N(N-1)/2] - M$.

Considering first the empty graph on $N$ vertices. Since there are no edges, there can be no loops, and the graph energy is
\begin{align}
E_{\emptyset} = Ne^{pv_0^2},
\end{align}
where we have used $g_V = 1$. We now examine the addition of a single edge. There is no preferred location for this edge, as every graph of one edge on $N$ vertices is isomorphic. The energy of all such graphs is
\begin{align}
E_{M =1} = (N-2)e^{pv_0^2} + 2e^{p(1-v_0)^2}.
\end{align}
As one more edge is created, there are two possible graph configurations: one in which the two edges are connected ($E_{\angle}$) and one in which they are separate ($E_{\parallel}$).
\begin{align} 
E_{\angle} &= (N-3)e^{pv_0^2} + 2e^{p(1-v_0)^2} + e^{p(2-v_0)^2},\\
E_{\parallel} &= (N-4)e^{pv_0^2} + 4e^{p(1-v_0)^2}.
\end{align}
Of these two configurations, separate edges are always favoured for any real $v_0$, and this condition is independent of both $N$ and $p$.

Repeating this process reveals that at each step the lowest energy graph with $M$ edges is one in which these edges are completely separate, so long as such separation is possible. Until loop formation becomes required, the lowest energy at each step is
\begin{align}
E = (N-2M)e^{pv_0^2} + 2Me^{p(1-v_0)^2}.
\end{align}
so that the rate of change of energy as edges are formed is
\begin{align} \label{deltaE}
\left.\delta E\right|_{M<N/2} = 2[e^{p(1-v_0)^2} - e^{pv_0^2}]
\end{align}
The maximum number of edges that can form before there are no longer any vertices of degree 0 is $M = N/2$. At this value of $M$, the graph is completely populated by disjoint edges and is 1-regular (every vertex has degree 1), with energy
\begin{align}
E = Ne^{p(1-v_0)^2}.
\end{align}
Further edge formation will not be able to continue the trend of disjoint edges forming. This will lead to a sudden change in the slope of the plot of energy against the number of edges. This discontinuity is general, and is seen whenever the maximum degree of the graph changes. Adding an edge that turns two vertices of degree 0 into vertices of degree 1 gives a change in energy given by eq.~\ref{deltaE}. However, once there are no vertices of degree 0 left, adding an edge that turns two vertices of degree 1 into vertices of degree 2 causes a change in energy
\begin{align}
\left.\delta E\right|_{N/2<M<N} = 2(e^{p(2-v_0)^2} - e^{p(1-v_0)^2})
\end{align}
with some corrections due to $H_L$ once more of these edges have formed.

After $M = N/2$ it becomes very difficult to proceed by insisting that the graph finds the lowest energy state before more edges form, as the number of possible graphs with $M$ edges becomes large and the formation of loops becomes possible. 

We now turn to the case where we start from the complete graph.  The complete graph with $N$ vertices has energy
\begin{align}
E_{K} = Ne^{p(N-1-v_0)^2}.
\end{align}
As with the empty case, all single edge deletions are equivalent, and this process may equivalently be considered as the formation of an edge hole. There are two distinct cases for the two-hole problem: where the two holes are connected ($E_{\angle^*}$), and where they are separate($E_{\parallel^*}$).
\begin{align}
\begin{split}
E_{\angle^*} &= (N-3)e^{p(N-1-v_0)^2} + 2e^{p(N-2-v_0)^2} \\
&+ e^{p(N-3-v_0)^2}, \\
E_{\parallel^*} &= (N-4)e^{p(N-1-v_0)^2} + 4e^{p(N-2-v_0)^2}. \\
\end{split}
\end{align}

We have neglected loops here for two reasons. Firstly, in most of the calculations of loops in quantum graphity presented here, the loops considered are shortest-path loops, also termed minimal loops.  As minimal loops best characterise a lattice structure , they are therefore important for yielding a crystalline ground-state graph. In the complete graph, shortest-path loops cannot form as there are no shortest paths of lengths greater than one (the distance between any two vertices is one). The second reason loops can be neglected here, which is independent of the loop-counting methods and definitions employed, is that the valence energy in the vicinity of the complete graph is expected to be many orders of magnitude greater than the loop energy. The number of loops of length $L$ that pass through single vertex $a$ is bounded from above by
\begin{align}
P(a,L) < cv^{L-1}
\end{align}
where $v$ is the degree of vertex $a$ and $c$ is some constant, the exact value of which is not important for this discussion. We see that the maximum possible contribution from the loops term is linear in the degree of each vertex, $v$, whereas the valence energy scales as $e^{v^2}$. Near the complete graph $v$ will be large and loops will therefore contribute negligibly to the energy.

As with the empty case, disjoint holes in the complete graph are most favourable. After $M^*<N/2$ disjoint holes have formed, the energy is
\begin{align}
E = (N-2M^*)e^{p(N-1-v_0)^2} + 2M^*e^{p(N-2-v_0)^2}.
\end{align}
When $M^* = N/2$, the graph is ($N/2$)-regular and experiences a sudden drop in energy, analogous to the empty case. Therefore every edge deletion near the complete graph changes the graph energy by
\begin{align}
\left.\delta E\right|_{M^*<N/2} = 2[e^{p(N-2-v_0)^2} - e^{p(N-1-v_0)^2}]. \label{eq:dE_complete}
\end{align}
Note that the energies associated with creation of edges from the empty graph and the creation of holes in the complete graph are different, implying egde-hole asymmetry. For large $N$, energies and rates of change of energy are much greater near the complete graph than near empty, due to the appearance of N in the exponent in eq.~\ref{eq:dE_complete}.  We discuss other manifestations of edge-hole asymmetry in this model, below.

\begin{figure}
\includegraphics[width=0.9\columnwidth,clip]{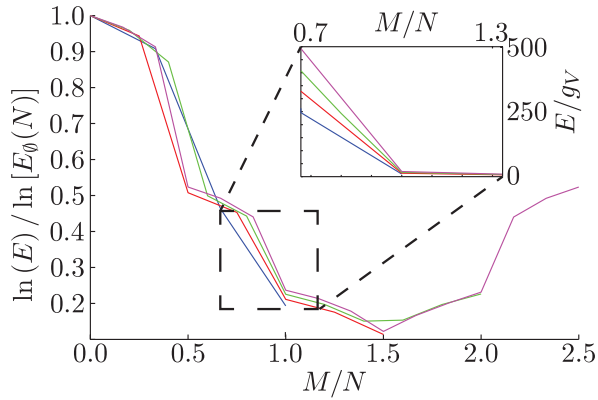}
\caption{\label{fig:ExactN6} Log to base e of the ground state energy of a graph on $N$ vertices normalised to the log of the empty graph energy, as a function of the number of edges $M$, calculated explicitly for up to N = 6. When N is large enough $(N-1)/2 > v_0/2$, the ground state is a 3-regular graph occurring at $M/N = v_0/2 = 3/2$. The plot consists of linear segments of different gradient. The apparent steep steps in the plot are due to the logarithmic energy scale, as shown in the inset. The inset shows a closeup of the energy divided by $g_V$ for $N=3, 4, 5,$ and $6$ vertices on a linear scale in the vicinity of $M/N = 1$.  In both figures $N=3$ is blue, $N=4$ is red, $N=5$ is green, and $N=6$ is magenta.} 
\end{figure}

To explore the energetics of the QG Hamiltonian, we also performed numerical simulations of the energy of finite models with up to 6 vertices, as a function of the number of edges (Fig.~\ref{fig:ExactN6}). Algorithmic loop counting was employed to give a more accurate calculation of the absolute ground state. Energy is plotted as a function of $m = M/N$. There are apparent steps in the energy plot, but these are just artefacts of the logarithmic scaling and the limited number of points plotted. A closer look given in  the inset of Fig.~\ref{fig:ExactN6} reveals that the energy plot consists of several linear segments of different gradient.

As expected, there is a discontinuity in energy each time the maximum valence of the graph drops, and the global minimum occurs when $m = 3/2$, which is the requirement for a 3-regular graph. We expect this result to be general, with the global minimum of any quantum graphity model occurring at $m = v_0/2$. For a finite graph, $v_0$-regularity is not always possible, as $M = v_0/{2N}$ may not be an integer, but this is not a problem for the $N\rightarrow\infty$ limit of the model, as a $k$-regular graph on infinite vertices is always possible.


In the $N\rightarrow\infty$ limit and when $m$ is far from $v_0/2$ so that the valence term dominates, the steps as the maximum valence of the ground state decreases will be seen at $m = n/2$ for any integer $n$. Closer to $m = v_0/2$ the loops term will become more important, and it is not clear exactly what effect this will have on the step structure. To study this behaviour numerically at larger N we apply the epitaxial approximation discussed below.

\subsection{Epitaxial approximation}
The number of possible graphs on $N$ vertices is $2^{N(N-1)/2}$. This number becomes large very quickly, so to find the lowest energy graph on $N$ vertices when $N\gtrsim 7$ we have imposed certain restrictions. One such restriction comes from assuming that geometric space in quantum graphity forms in a manner analogous to the epitaxial growth of crystals, rather than considering unconstrained optimisation of the graph at each step.

To explore the epitaxial growth of spatial `grains', we start with a fixed, frozen configuration, to which we add or delete one edge. Treating this new edge or hole as a particle, we allow the edge to explore the graph until it finds a (possibly degenerate) local minimum energy configuration. We then repeat for the next edge or hole. In this way, rather than calculating the energy of every possible graph on $N$ vertices with $M$ edges, we  only consider those graphs attainable by adding or deleting a single edge to some fixed initial graph. This approximation, although physically motivated, is not guaranteed to find the true minimum configuration for fixed number of edges, but it does provide insight into how it is likely that spatial domains could develop under the assumptions of QG.

Another approximation required to perform calculations for large $N$ is to truncate loop counting at loops of length 7. This approximation allows the use of explicit formulas for calculating the number of loops in a graph from the trace of the adjacency matrix, and such formulas are not known for loops of length $>7$.  The lack of contribution from longer loops means that the graph does not benefit energetically by forming an extended lattice. Rather, there is a preference for the formation of small disjoint subgraphs. 

\begin{figure*}
\centering
\includegraphics[width=1\linewidth]{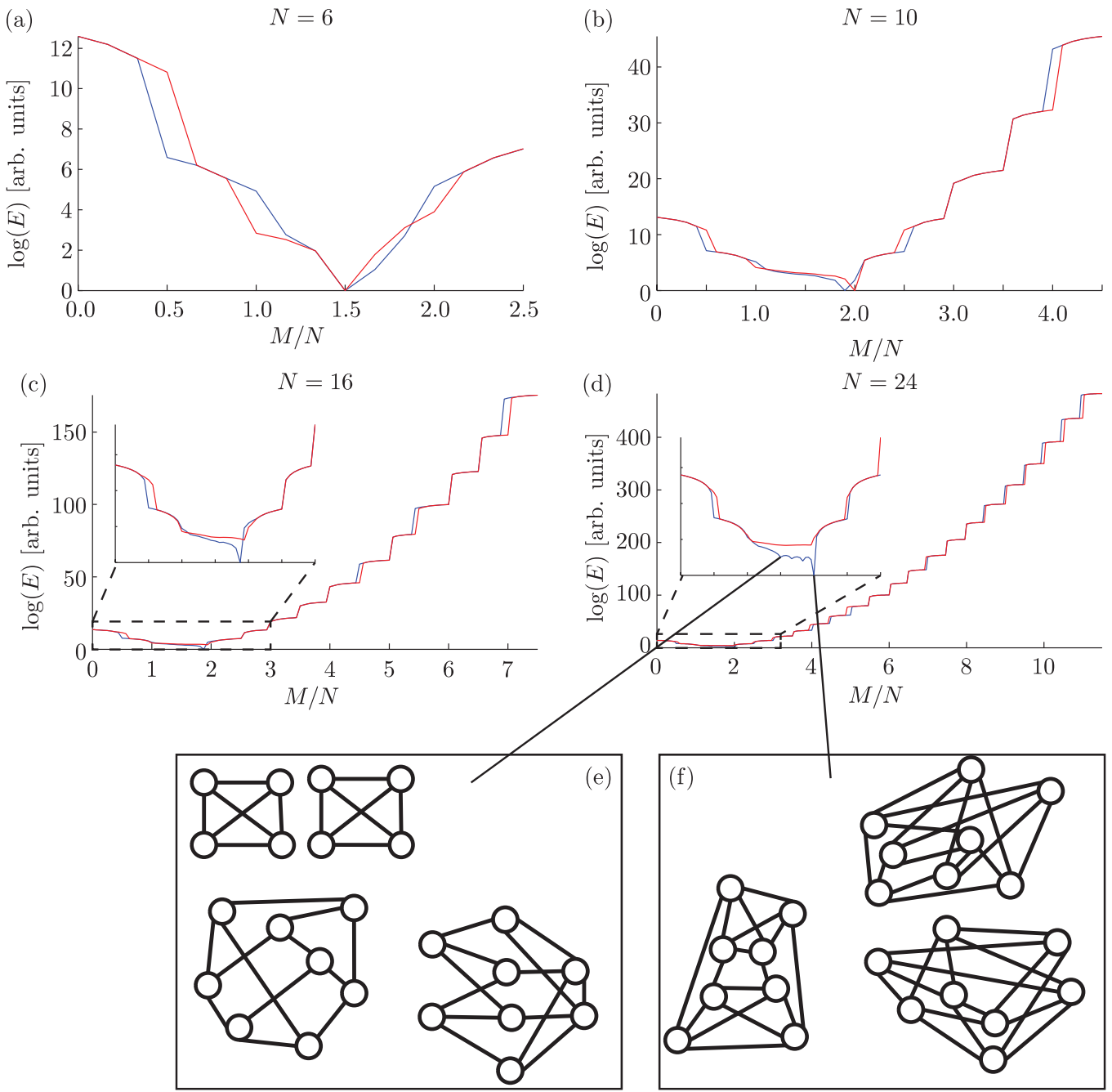}
\caption{\label{fig:EpitaxyEnergyN24} Full energetics of the graph on up to 24 vertices in the epitaxial approximation. Both complete (red) and empty (blue) starting conditions were used, and give quite different results. This indicates a hysteretic effect may be present, and explicitly shows the particle-hole asymmetry of the model. Two degree-regular graphs are reached under the epitaxial approximation starting from an empty graph on 24 vertices. (e) Shows the 3-regular graph at $M/N = 3/2$, and (f) shows the 4-regular graph at $M/N = 2$. Both graphs consist of several disjoint subgraphs. The fact that the 3-regular graphs contains more disjoint subgraphs may account for its higher energy, despite the fact that a 3-regular graph minimizes $H_V$.}
\label{fig:Epitaxialenergyplot}
\end{figure*}

The energetics of epitaxial growth of the quantum graphity model were calculated for all cases up to $N = 24$ starting from both the empty (edge-addition) and the complete graphs (hole-addition) (see Fig.~\ref{fig:EpitaxyEnergyN24}). One of the most immediately striking effects visible is the asymmetry of the energy from each starting point.  The steps associated with a drop in maximum valence that were predicted in the case where the graph reaches a ground state for each fixed value of $M$ are seen here, although we can see that the exact location of these steps differs between the complete and empty initial state approaches. This leads to an apparent hysteresis.

Looking more closely at the region where the absolute ground state of the system is expected to lie, the asymmetry of the edge- and hole-addition approaches is even more clear. Most importantly, the lowest energy state found in this approach is not a 3-regular graph, but rather a 4-regular graph. To examine why this may be the case, the particular configurations of the graph are drawn at both the point where we except the ground state to occur ($m = 3/2$, figure~\ref{fig:EpitaxyEnergyN24} (e) ) and the point at which we actually see the lowest energy ($m = 2$, figure~\ref{fig:EpitaxyEnergyN24} (f) ).

In both cases, the graph is composed of several smaller disjoint subgraphs. It is unclear if there is an artefact of the epitaxial approximation, or a more general feature of the QG Hamiltonian. The requirement of truncation of the sum over all loop lengths in $H_L$ may mean that the Hamiltonian is minimised when the graph consists of disjoint subgraphs of length $L_{\max}$, as there will be no energetic benefit of forming domains larger than this and the graph grains additional loops which wrap around the entire grain. This may indicate that an additional constraint is required in the QG Hamiltonian to ensure that the absolute ground state of the system is connected, as a graph consisting of small disjoint subgraphs cannot represent space as we experience it.

The fact that the lowest energy state is not 3-regular is not expected to be general, and may be a result of either the epitaxial approximation or the small size of the graph. The 3-regular graph in  Fig.~\ref{fig:EpitaxyEnergyN24} (e) consists of 4 disjoint subgraphs, whereas the 4-regular graph in Fig.~\ref{fig:EpitaxyEnergyN24} (f) is composed of only 3 disjoint subgraphs. The 3-regular graph contains two subgraphs of only 4 vertices, making the formation of 6-loops impossible (recall that these are the loops we have chosen to contribute most significantly to the loop energy). This is likely to be the main reason why the 4-regular graph is a lower energy than the 3-regular graph. In the early stages of epitaxial growth, the graph has been primarily concerned with minimising the number of vertices that vary from ideal valence by a significant amount (here a $|v_i - v_0|$ of 2 or 3 may be considered significant). In doing so, the graph has become stuck in a configuration that cannot form the 6-loops or lattice-like structure that would minimize the energy.

\section{Conclusion}
Quantum graphity is a background independent model in which space is represented by a dynamical graph and properties such as geometry and locality are emergent at low energies. We explored the way this model proceeds from a high-energy pre-geometric starting point towards a low-energy geometric state as edges of the graph are added or deleted. To simulate larger graphs, we made use of an epitaxial approximation as well as an explicit method for counting short loops. Values for the rate of change of the energy of the graph as edges are added or deleted close to the high-energy end points were obtained analytically, and the full energy curve as a function of number of edges was plotted numerically up to $N = 24$. Future work may be able to compare these features of the model with observations of the early universe in order to fix values for the parameters of the model, making it potentially falsifiable. For example, a Fermi's golden rule argument may make it possible to determine the rate of formation of spatial domains, given the differences in energy between two graphs.

Evolution towards the ground state is qualitatively different between the two possible starting points: the complete graph and the empty graph. The complete graph on $N$ vertices has a metric dimension of $N-1$, so a quantum graphity universe starting from the complete graphs begins with an effectively infinite-dimensional space that unfolds into the 3-dimensional space we see today. In this approach the early universe is highly connected, offering a possible solution to the horizon problem of cosmology, and under the epitaxial approximation space remains disconnected until it reaches an energetic minimum, which is in agreement with our current understanding of our universe. This may indicate the need to introduce addition constraints to the model in order to impose connectivity of the system (see \cite{bib:ChenPlot2013}, for an example of a graph model that imposes such a constraint).

On the other hand, starting from the empty graph we initially have no edges at all, which is a stronger notion of spacelessness and corresponds more closely to the notion of a universe evolving from ``nothing". Rather than an unfolding of space as we see with the complete graph starting point, starting from the empty graphs leads to the formation of disconnected spatial graphs which later stitch together. Whether the resulting ground state of the model is connected or not likely depends strongly on assumptions of how the graph evolves, and under the epitaxial approximation for $N = 24$ we have seen that the lowest energy state is indeed one consisting of several disjoint subgraphs. A disconnected final state would be a poor representation of our universe, unless each individual grain is large enough to represent a universe in itself. In this case we have a multiverse-type picture emerge, in which there is no adjacency between the distinct universes. However, if edges of the graph are relatively free even in the low energy state, connections between previously disconnected universes may form spontaneously. 

Many discussions of quantum graphity, including this one, have assumed a lattice-like ground state to be the ideal low-energy geometric state. It is not quite clear if this assumption is justified, as the honeycomb lattice has only been shown to be metastable, and finding the absolute ground state explicitly and exactly may not be possible. By analogy with the lattice structures of crystals, a lattice-like ground state may lead to an inherent anisotropy, similar to the birefringence exhibited by regular crystal lattices. Furthermore, the assumption of a regular lattice for the ground state forces the Hamiltonian parameter $v_0$ to be an integer, where there is no other justification for this. Hence crystalline order is not the only possibility. Non-integer values of $v_0$ would lead to glassy structures forming the ground state graphs of the model, where in the $N\rightarrow \infty$ limit the mean valence of all of the vertices will be $v_0$. Such a graph may be more isotropic than a regular lattice graph, which is a desirable feature for a model of emergent geometry as it reflects what is observed in the current universe.

Future work would consider the explicit time sequence of how geometrogenesis can occur and especially the epitaxial growth of geometric space.

\section{Acknowledgements}
ADG acknowledges the ARC for financial support (DP130104381). We would like to thank Jared H. Cole for helpful comments and conversations.

\end{document}